\documentclass[twocolumn,showpacs,preprintnumbers,amsmath,amssymb]{revtex4}

\usepackage{graphicx}
\usepackage{dcolumn}
\usepackage{bm}

\begin{document}


\bibliographystyle{prsty}
\input epsf

\title{Unusual magnetic properties of the low-dimensional quantum magnet Na$_{2}$V$_{3}$O$_{7}$}

\author{J. L. Gavilano$^{1}$, E. Felder$^{1}$, D. Rau$^{1}$, H. R.
Ott$^{1}$, P. Millet$^{2}$, F. Mila$^{3}$, T. Cichorek$^{4}$ and A. C.
Mota$^{4}$}

\affiliation{ $^{1}$ Laboratorium f\"{u}r Festk\"{o}rperphysik,
ETH-H\"{o}nggerberg, CH-8093~Z\"{u}rich, Switzerland \\
$^{2}$ Centre d'Elaboration des Mat$\acute{e}$riaux et d'Etudes
Structurales, 29, rue J. Marvig, 31055 Toulouse Cedex, France \\
$^{3}$ Institute of Theoretical Physics, Federal Institute of Technology Lausanne,  CH - 1015 Lausanne, Switzerland\\
$4$ Max-Planck-Institut f\"{u}r Chemische Physik fester Stoffe,
01187 Dresden, Germany}

\date{\today}

\begin{abstract}
We report the results of low-temperature measurements of the specific heat $C_{p}(T)$, $ac$ susceptibility $\chi_{ac}(T)$ and $^{23}$Na nuclear magnetic resonance NMR of Na$_{2}$V$_{3}$O$_{7}$.  At liquid He temperatures $C_{p}(T)/T$ exhibits broad field-dependent maxima centered at $T_{0}(H)$, which shift to higher temperatures upon increasing the applied magnetic field $H$.  Below 1.5 K the $ac$ magnetic susceptibility $\chi_{ac}(T)$ follows a Curie-Weiss law and exhibits a cusp at 0.086 mK which indicates a phase transition at very low temperatures.  These results support the previous conjecture that Na$_{2}$V$_{3}$O$_{7}$ is close to a quantum critical point (QCP) at $\mu_{0}H \approx 0$ T. The entire data set, including results of measurements of the NMR spin-lattice relaxation $T_{1}^{-1}(T)$, reveals a complex magnetic behavior at low temperatures. We argue that it is due to a distribution of singlet-triplet energy gaps of dimerized V moments. The dimerization process evolves over a rather broad temperature range around and below 100 K. At the lowest temperatures the magnetic properties are dominated by the response of only a minor fraction of the V moments.
\end{abstract}
\pacs{
75.40.-s, 75.30.kz, 76.60.-k, 75.30.Et,
}
\maketitle


\section{INTRODUCTION}

Low-dimensional quantum magnets have recently been in the focus of both experimental and theoretical research\cite{Millet98,Ueda98,Korotin99}.  Among different relevant materials of interest, oxides played a major role, in particular  a variety of vanadium oxides. A few years ago another member, Na$_{2}$V$_{3}$O$_{7}$, was added to this family by P. Millet and coworkers.  The arrangement of the V magnetic moments in this compound exhibits very peculiar geometrical features\cite{Millet99} and in some respects, this material may be regarded as a spin $S = 1/2$ ladder-type compound with periodicity along the rung direction.  This periodicity affects the topology of the arrangement of the magnetic moments.  From the theoretical point of view it is still unclear  to what extent this affects the magnetic properties of the system\cite{Schulz96,Kawano97}.

Recent work has revealed that Na$_{2}$V$_{3}$O$_{7}$ is an insulator with unusual low temperature properties\cite{Gavilano2002,Gavilano2003}.  Considering the temperature dependence of the dc-magnetic susceptibility $\chi(T)$, it was argued\cite{Gavilano2002,Gavilano2003} that at temperatures $T > 100$ K, all the V ions adopt a tetravalent configuration ($S = 1/2$). The resulting moments act paramagnetically but are exposed to substantial antiferromagnetic interactions.  If we denote the ratio between an effective number of uncompensated V magnetic moments, each of spin $S = 1/2$, and  the number of all the V ions as  $n_{mag}$, it follows that for $T > 100 K$, $n_{mag}$ = 1, and the paramagnetic Curie temperature $\Theta_{p}$ is of the order of -200 K. Below 100 K, the V moments appear to be gradually quenched, such that below 20 K, only one out of nine V moments, exposed to a much weakened mutual coupling, contributes to $\chi(T)$. In this regime, $n_{mag}$ = 1/9.  From the $^{23}$Na-NMR response of Na$_{2}$V$_{3}$O$_{7}$ it was concluded\cite{Gavilano2003} that in the presence of modest external fields $H$, up to 7 T, and at temperatures of the order of 2.5 K and below, phase transitions occur at field dependent critical temperatures $T_{a}$.  Since $T_{a}$ shifts towards zero upon reducing $H$, it was suggested that Na$_{2}$V$_{3}$O$_{7}$ is close to a quantum critical point (QCP) at $\mu_{0}H \approx 0$ T.

Na$_{2}$V$_{3}$O$_{7}$ crystallizes in a trigonal structure with $a = 10.886$ \AA $\:$ and $c = 9.5380$ \AA\cite{Millet99}.  The space group is $P31c$ and the unit cell contains 6 formula units. The V ions  occupy the geometric centers of O$_{5}$ square pyramids, which in turn form the walls of nanotube-type structural elements with a diameter of 5 \AA. These tubes are aligned along the $c-$axis of the crystal structure and the V ions occupy 9 sites on individual rings on the circumference  of the tubes.   The interactions among the V moments vary over a wide range\cite{Mazurenko2004}.  Generally, they involve some hybridization between V-neighboring O orbitals.  The Na ions occupy four inequivalent sites.  One of them, Na1, is located in the center of  the nanotubes, and the rest, Na2, Na3, and Na4, is arranged around them.  All the Na ions are bonded to oxygen ions which are at the vertices of complicated polyhedra around the Na ions.  The same oxygen ions also form O$_{5}$ square pyramids around the V ions.

We report on measurements of the specific heat $C_{p}(T)$, ac-magnetic susceptibility $\chi_{ac}(T)$ and $^{23}$Na-NMR on Na$_{2}$V$_{3}$O$_{7}$.  Our low-temperature $C_{p}(T)/T$ and $C_{p}(T)$ data reveal broad maxima at field dependent temperatures. Below 1.5 K, $\chi_{ac}(T)$ increases in a Curie-Weiss type fashion with decreasing temperature.  A narrow maximum at 0.086K indicates a phase transition.  Our new results support our previous claim that Na$_{2}$V$_{3}$O$_{7}$ is close to a quantum critical point QCP at $H = 0$ T. Anomalies in the temperature and field dependencies of the NMR response and $\chi(T)$ at low temperatures indicate different regions in the $H-T$ phase diagram with different magnetic properties. It is argued that this reflects a complex distribution of singlet-triplet energy gaps of dimerized V moments.

\section{EXPERIMENTAL DETAILS}

\subsection{ Measurements}

The $C_{p}(T)$ measurements were made using a relaxation-type method at temperatures between 0.3 and 40 K in either a $^{3}$He refrigerator, or a $^{4}$He-flow cryostat, both fitting into the bore of a 7 T superconducting magnet.  For these measurements the sample consisted of small crystals, densely packed into a small thin-walled Cu container. In order to gain access to $C_{p}(T)$ of Na$_{2}$V$_{3}$O$_{7}$ alone, the specific heat of the empty Cu container was measured separately.  For the measurements of the ac-susceptibility between 0.018 and 1.8 K, we used an ac-impedance bridge with an rf-SQUID (superconducting quantum interference device) as the null detector. In this custom-built arrangement, the detection coils and the sample are placed inside the mixing chamber of a dilution refrigerator, in direct contact with the liquid helium mixture.  For our measurements the sample was kept stationary inside the gradiometer-type arrangement of the detection coils\cite{Dumont2002}.  The ac driving field had an amplitude of 1.3 mOe and a frequency of 80 s$^{-1}$.  

The $^{23}$Na NMR measurements between 0.1 and 300 K, were performed using standard spin-echo techniques and home-built
spectrometers and probes.  An unsuccessful attempt was made to observe the $^{51}$V NMR signal at high (295 K) and low (near 1 K)
temperatures.  We concluded that at high temperatures, the spin-spin relaxation time $T_{2}$ for the $^{51}$V nuclei must be
very short, less than 5 $\mu s$.  At low temperatures, the $^{51}$V NMR signal may be distributed over a very broad range of
frequencies. The central Zeeman transition $(+1/2 \leftrightarrow -1/2)$ of the $^{23}$Na NMR spectra was obtained by integrating
the spin-echo signal after a $\pi/2-\pi$, $rf$-pulse sequence, at a fixed applied magnetic field by changing stepwise the frequency and also at a fixed frequency by changing stepwise the magnetic field.  The quadrupolar wings could not be observed except, perhaps, at very low-temperatures, where a weak NMR signal intensity was recorded around the central transition.  Indeed, the results of a calculation of electric field-gradients at the Na sites, using a point-charge model, confirm that for our randomly oriented powder-sample, the NMR signal intensity from the quadupolar wings is distributed over a broad frequency range. Therefore, all the NMR results presented below refer only to the central transition of the $^{23}$Na spectrum.  The spin-lattice relaxation time $T_{1}$ was measured by first destroying the
nuclear magnetization upon the application of a long comb of $rf$ pulses, waiting for a variable recovery time $t$, and then
measuring the integrated spin-echo intensity after a $\pi/2-\pi$, $rf$-pulse sequence.  For the NMR measurements below 20 K the sample was in contact with the $^{3}$He-$^{4}$He mixture of a dilution refrigerator.  At temperatures above 20 K, a $^{4}$He flow cryostat was used to cover this temperature regime.  Part of the NMR data were  published previously in Ref. 8.

\subsection{Samples}

The material was prepared by growing very small crystals from melts of
the starting composition Na$_{1.9}$V$_{2}$O$_{5}$\cite{Millet99} at
reduced pressure.  Densely packed material was used for measurements
of $C_{p}(T)$ and $\chi_{ac}(T)$.  For the NMR experiments the sample
consisted of loosely packed powder.

\section{EXPERIMENTAL RESULTS AND RELATED CALCULATIONS}

\subsection{Specific heat}


\begin{figure}
\includegraphics[width=0.8\linewidth]{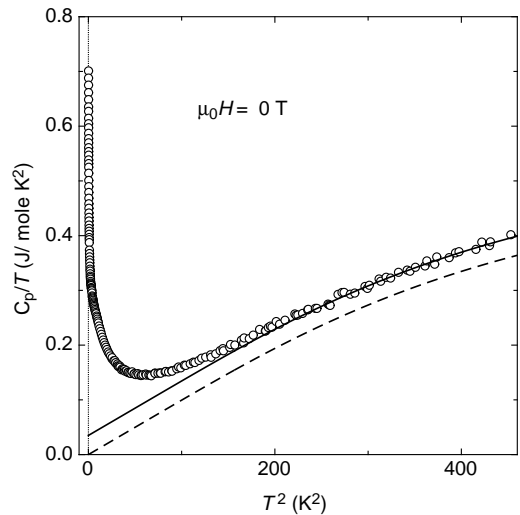} 
\caption{\label{fig1:epsart4}
  $C_{p}/T$ as a function of $T^{2}$ for
  Na$_{2}$V$_{3}$O$_{7}$ at zero applied magnetic field.
  The solid line represents the sum expressed in Eq.  1 with the
  parameters quoted in the text. The broken line represents the lattice contribution.
   }
\end{figure}
In Fig.  1 we display $C_{p}/T$ as a function of $T^{2}$, measured in zero applied magnetic field.  Above 15 K the $C_{p}$ data is well represented by
\begin{equation}
    C_{p} = \gamma T +  C_{DE} ,
    \label{eq:1}
\end{equation}
with $\gamma = 35$ mJ/(mol K$^{2}$) and $ C_{DE}$, emphasized by the dotted line, representing the specific heat of the lattice due to acoustic phonons (Debye model with $\Theta_{D}$ = 125 K) and the lowest optical mode (Einstein model) at 88 cm$^{-1}$\cite{choi2003}.  The temperature variation  according to Eq. 1 is represented by the solid line in Fig.  1.  At temperatures above 15 K, the specific heat is practically independent of the magnetic field $H$, but at lower temperatures it varies significantly with increasing $H$.

Considering the specific heat $C_{m}(T) \equiv C_{p}(T) - C_{DE} $, $i.e.$, in excess of the usual lattice contribution of insulators, the parameter $\gamma$ turns out to be rather large.  Since Na$_{2}$V$_{3}$O$_{7}$ is an insulator, the corresponding contribution to $C_{m}(T)$ is definitely not of (itinerant) electronic origin but most likely due to magnetic degrees of freedom. In Figs.  2 and 3 we display the excess specific heat $C_{m}$, as defined above, plotted as  $C_{m}/T$ versus $T$, including data for external magnetic fields  ranging from 0 to 7 T. The temperature- and field-independent contribution registered above $T$ = 15 K corresponds to the large $\gamma T$ term in $C_{p}(T)$ mentioned above.

\begin{figure}
\includegraphics[width=0.8\linewidth]{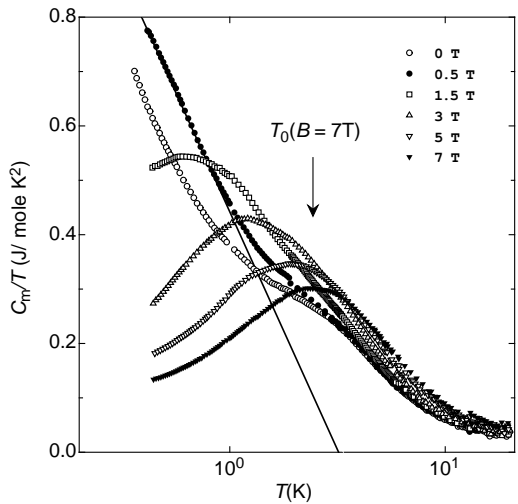} 
\caption{\label{fig2:epsart4}
$C_{m}/T \equiv (C_{p} - C_{DE})/T $ as a function of ln$T$ for  Na$_{2}$V$_{3}$O$_{7}$ measured in different external magnetic
  fields.  The solid line emphasizes the logarithmic temperature variation.
  }
\end{figure}
In an applied magnetic field of 0.5 T we observe a logarithmic divergence at very low temperatures, emphasized by the solid line in Fig.  2, which represents
\begin{equation}
     C_{m}(T)/T =  -0.38 \cdot \ln (T/T_{K}) ,
    \label{eq:2}
\end{equation}
where $C_{m}$ is measured in units of  J mol$^{-1}$K$^{-2}$ and $T_{K} =  3.2$ K. This behavior is identified between 0.37 K, the lower end of the covered temperature range, and 0.83 K. A similar dependence, with the same pre-factor  but a lower $T_{K}$  characterizes the $ C_{m}(T)/T $ data measured in zero field. Logarithmic divergences of $C_{m}/T$ at low temperatures were often reported for a variety of fermionic systems close to QCPs\cite{Stewart2001,vonLohneysen1999,Cox1998}.

\begin{figure}
\includegraphics[width=0.8\linewidth]{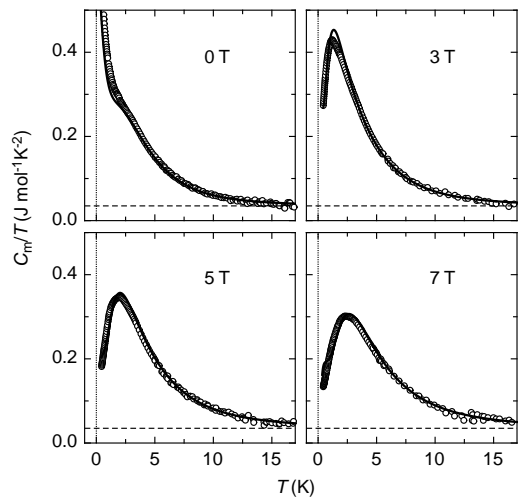} 
\caption{\label{fig3:epsart4}
$C_{m}/T  $ as a function of $T$ for  Na$_{2}$V$_{3}$O$_{7}$ measured in different external magnetic
  fields.  The solid lines represent the best fits to the data using a model described in the text. The horizontal broken lines represent the constant  $\gamma$ term (see Eq. 1).  }
\end{figure}
The $C_{m}(T)$ data, obtained for fields exceeding 1 T, however, exhibit field dependent maxima in $C_{m}(T)/T$ and, even for $H=0$ T,  in $C_{m}(T)$, as demonstrated in Figs. 2, 3 and 4. The temperatures of the maximum for $C_{m}(T)$ does not vary much upon changing the magnetic field $H$, at least for $\mu_{0} H  <  6$ T. The position of the maximum of $C_{m}(T)/T$,  however, shifts to higher temperatures with increasing field (see Fig. 5). The solid lines in Figs. 3 and 4 represent the best fits to the data using a model to be described below.  The model is chosen such that the calculation of   $C_{m}(T)$ and $C_{m}(T)/T$ provides an acceptable agreement with experiment.  The data below 20 K (see Figs. 1, 3 and 4) suggest that  $C_{m}(T) = \gamma T + C_{st}(T)$, whereby $C_{st}(T)$ contributes significantly only below 15 K. In the following we concentrate on $C_{st}(T)$ and the $\gamma T$ term will only be considered again in the discussion section.
\begin{figure}
\includegraphics[width=0.8\linewidth]{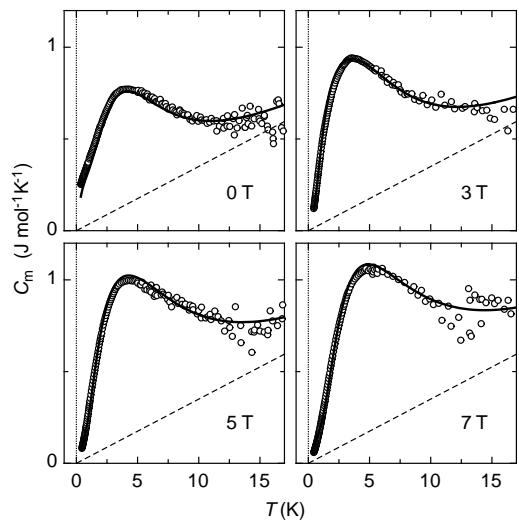} 
\caption{\label{fig4:epsart4}
$C_{m} $ as a function of $T$ for  Na$_{2}$V$_{3}$O$_{7}$ measured in different external magnetic
  fields.  The solid lines represent the best fits to the data using a model described in the text. The broken lines represent  the $\gamma T$ term (see Eq. 1).  }
\end{figure}
 
\begin{figure}
\includegraphics[width=0.8\linewidth]{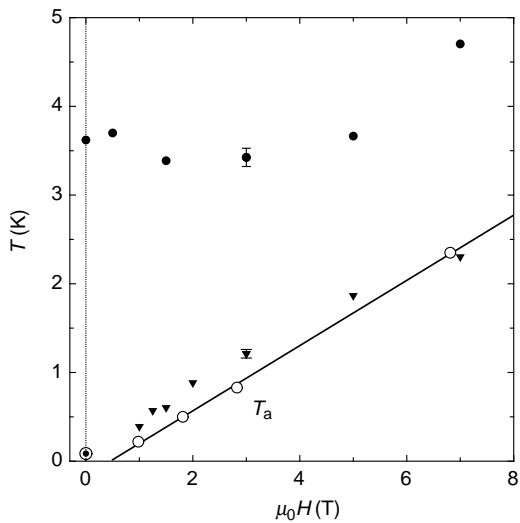} 
\caption{\label{fig5:epsart4}
$T_{max}$, the position of the maxima of $(C_{m}(T) - \gamma T)$ (filled circles) and $C_{m}(T)/T$ (filled triangles), as a function of $H$ for  Na$_{2}$V$_{3}$O$_{7}$.  The circle with the center dot represents the sharp peak in $\chi_{ac}(T)$ (see Fig. 8) and the empty circles the position of the maxima at  $T_{a}$ in the NMR spin-lattice relaxation rate.
}
\end{figure}

It turned out that the best agreement between experiment and model calculation could be obtained by assuming that $C_{st}(T)$ arises from excitations in a collection of dimerized pairs of V moments, $i.e.$, an ensemble of singlet-triplet units with a distribution $P(\Delta)$ of singlet-triplet energy gaps $\Delta$. A more detailed account on the justification of this choice is given  in the discussion section. 
We assume that  $P(\Delta)$ is the sum of two Gaussians,  $P_{1}(\Delta)$ and $P_{2}(\Delta)$. $P_{1}(\Delta)$ is narrow
with a width $W_{1}/k_{B}$ of the order of 4 K ($HWHM = 0.83 \times W_{1}/k_{B} = 3.2$ K)  centered at $T_{max1} = 0$ K and  $P_{2}(\Delta)$ is a broader Gaussian centered at $T_{max2} \approx  9$  K with a width  $W_{2}/k_{B}$ of about 10 K ($HWHM = 0.83 \times W_{2}/k_{B} = 8.$3 K).  The distribution of $P(\Delta)$ is thus given by
\begin{eqnarray}
    P(\Delta) &=& \frac{0.055}{W_{1} \sqrt(\pi)} \exp \Big\{-\left[  \frac{\Delta- k_{B}T_{max1}}{W_{1}} \right]^{2}\Big\} 
    \nonumber \\  &+&\frac{0.11}{W_{2} \sqrt(\pi)} \exp\Big\{-\left[ \frac{\Delta - k_{B}T_{max2}}{W_{2}} \right]^{2} \Big\}.
    \label{eq:3}
\end{eqnarray}

Keeping in mind that a formula unit of the substance contains 3 V ions or 3/2 V pairs, and that at low temperatures, $n_{mag} = 1/9$ as explained in the introduction, the prefactors are chosen to fix the total population (area) to $(1/9)(3/2) = 1/6$ pairs per formula unit. The ratio between the relative populations of the narrow and the broad Gaussian is about  0.5 and is the same for all the data sets. Taking $W_{1}$, $ T_{max2}$ and $W_{2}$ as free parameters  ($ T_{max1}$ is fixed to zero) varying slightly with $H$, we can reproduce very well all the available $C_{m}(T)$ and $C_{m}(T)/T$ data as demonstrated by the solid lines in Figs. 3 and 4. For this we have calculated the specific heat due to a collection of individual singlet-triplet units in an applied magnetic field $H$. The levels are considered to be $\delta$-peaks with the energy of the singlet level at zero and  of the  triplet levels at $\Delta$, $\Delta+2\mu_{B}H$ and $\Delta-2\mu_{B}H$, respectively. The calculations are based on the distribution function $P(\Delta)$ given in Eq. 3.

\begin{figure}
\includegraphics[width=0.8\linewidth]{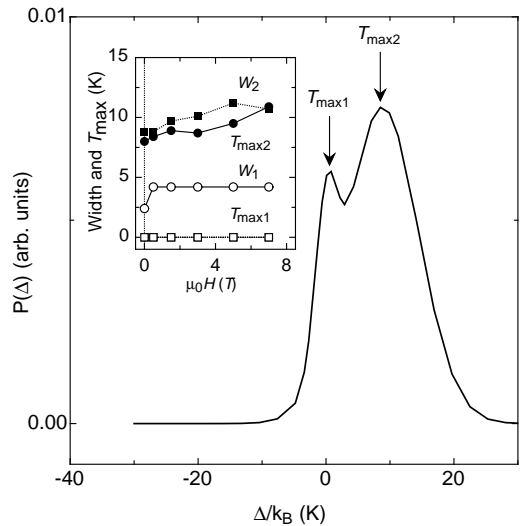} 
\caption{\label{fig6:epsart4}
Probability distribution $P(\Delta) = P_{1}(\Delta) + P_{2}(\Delta)$ of the spin singlet-triplet energy gap $ \Delta$. $P(\Delta)$ was obtained as the sum of two Gaussians from fitting the $C_{m}(T)$ and $C_{m}(T)/T$ data. Inset: Magnetic field dependence of the centers ($T_{max1}$ and $T_{max2}$)  and of the widths  ($W_{1}$ and $W_{2}$) of the Gaussians ($P_{1}$ and $P_{2}$).}
\end{figure}

$P(\Delta)$ in zero magnetic field, given by Eq. 3  and resulting from inserting the best fit parameters, is represented in Fig. 6. The inset displays the magnetic-field induced variations of the parameters  $W_{1}$, $ T_{max2}$ and $W_{2}$.  Ideally they should be field independent, but  the best agreement with experiment is obtained by invoking a weak field dependence which, most likely, reflects the limits of our model. Inspection of Fig. 6 reveals a small fraction of singlet-triplet units with negative gaps, $\Delta < 0$, implying that for a minority of pairs the spin-triplet configuration is energetically more favorable. This would be rather surprising if the inter-ring exchange was a simple superexchange between appropriate $d_{xy}$ orbitals, as assumed for instance in Ref. \cite{Saha-Dasgupta}, because the couplings would then all be antiferromagnetic.  But since Mazurenko and coworkers\cite{Mazurenko2004}, on the basis of an LDA+U estimate of the exchange integrals, conclude that inter-ring couplings are ferromagnetic, the quoted implication is actually plausible. The model accounts for the field-dependent maxima of both $C_{p}(T)$ and  $C_{p}(T)/T$ and for the substantial increase of  $C_{p}(T)/T$ with decreasing temperatures in external magnetic fields of 0 and 0.5 T, respectively (see Figs. 2 and 3).

The above discussion implies that the anomalies  of $C_{p}(T)$ and $C_{p}(T)/T$ do not reflect cooperative phase transitions.  However, the low-temperature anomalies at $T_{a}$ in the NMR spin-lattice relaxation $T_{1}^{-1}(T)$, reported in Ref. 8, reveal changes in the dynamics of the V moments in the same temperature range. We address this issue in the discussion section. Since the maxima in $C_{p}(T)/T$ or in $C_{p}(T)$ are well accounted for by our model and no other specific-heat anomalies are discernible, also the features in $T_{1}^{-1}(T)$ at $T_{a}$ cannot reflect transitions to conventional magnetically-ordered phases. 
\begin{figure}
\includegraphics[width=0.8\linewidth]{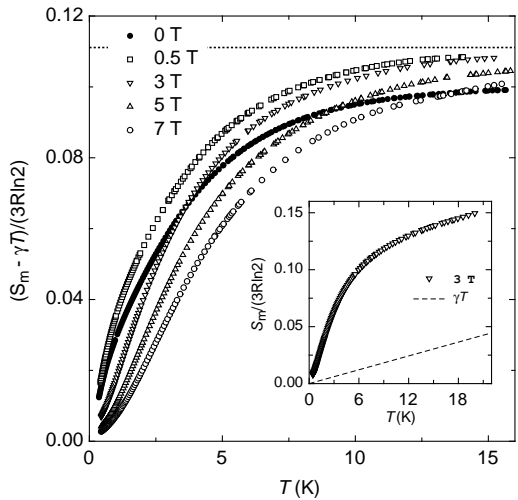} 
\caption{\label{fig7:epsart4}
Normalised magnetic entropy $(S_{m} - \gamma T)$ per mol of V ions as a function
of temperature for different external magnetic fields.  The horizontal broken lines represents $1/9$ of the normalized total entropy, $S_{total}/(3R \ln 2$) = 1. Inset: Temperature dependence of the magnetic entropy including the contribution of the $\gamma T$ term for $\mu_{0}H = 3 T$.
}
\end{figure}

The magnetic entropy $S_{m}(T)$ was obtained by numerical integration of $C_{m}(T)/T$ and asserting that $S_{m}(0) = 0$.  In Fig.  7 we display $S_{m}(T) - \gamma T$ for magnetic fields of 0, 0.5, 3, 5 and 7 T.  These data imply that only a very small magnetic entropy $S_{m}(T)- \gamma T$ is released at low temperatures, reaching  approximately $1/9$ of $3R \cdot \ln 2$ at 20 K  or above. This is  the value expected for the entropy of a system of dimers that is described by Eq. 3, and it implies that indeed, $n_{mag}$ = 1/9  below 20 K. This observation adds convincing support for our previous conjecture that the V magnetic moments in Na$_{2}$V$_{3}$O$_{7}$ are gradually quenched with decreasing temperature between 100 and 20 K.  As may be seen in the inset of  Fig.  7, above 9 K, $S_{m}(T)$ increases almost linearly with $T$,  reflecting the importance of the large $\gamma T$ term of the specific heat. As mentioned before, the significance of this term is addressed in the discussion section.

\subsection{ac-susceptibility}

\begin{figure}
\includegraphics[width=0.8\linewidth]{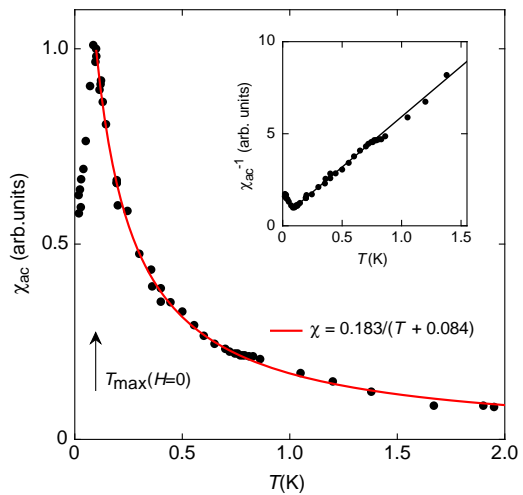} 
\caption{\label{fig8:epsart4}
Temperature dependence of the ac-magnetic susceptibility of
Na$_{2}$V$_{3}$O$_{7}$ in zero applied magnetic field.  The solid line
represents the best fit to a Curie-Weiss law.  The sharp
peak in $\chi_{ac}(T)$ at 0.086 K reveals a phase transition at an
unexpectedly low temperature. Inset: $\chi_{ac}^{-1}$ vs. $T$. }
\end{figure}
In Fig.  8 we display  the results for $\chi_{ac}(T)$, after subtraction of a $T-$independent signal.  Between 0.11 and 1.5 K the data is well
represented by
\begin{equation}
     \chi_{ac}(T) = \chi_{0}  + C/(T-\Theta_{p}) .
    \label{eq:4}
\end{equation}
The  very small paramagnetic Curie temperature $\Theta_{p} = -0.084$ K reveals a very weak coupling of the remaining magnetic
moments of V (see also the inset of Fig. 8).

Next we estimate the concentration of the V magnetic moments that contribute to  $\chi(T)$. Since the absolute value of $\chi_{ac}$ cannot be extracted directly from the $\chi_{ac}$ measurements, we simply match the Curie-Weiss parts of $\chi_{ac}(T)$ and $\chi_{dc}$ (see Ref.  8) at $T = 2$ K. From this procedure we find that the low-temperature Curie constant $C$ of Eq. 4  is only half of the corresponding value at temperatures above 2 K. This is interpreted as reflecting a further reduction of $n_{mag}$ by the same factor. In summary, from all the results for $\chi(T)$ we conclude that in Na$_{2}$V$_{3}$O$_{7}$ near room temperature, all the V moments are active and  $n_{mag} = 1$. Between 2 and 20 K, $n_{mag} = 1/9$ and between 0.11 and 1.5 K, $n_{mag} = 1/18$.

The very sharp peak in $\chi_{ac}(T)$ at $T_{max}(H \approx 0)$ = 0.086 K signals a phase transition involving spin degrees of freedom.  Other than a slight suppression of the amplitude of the peak, no field-induced changes are observed in $\chi_{ac}(T)$ measured in small fields of the order of 150 Oe (data not shown).  Since at these temperatures the concentration of active magnetic moments is very small ($n_{mag} = 1/18$), it is again very unlikely that this transition is of conventional ferro- or antiferromagnetic type.  Most probably it reflects the freezing of the few remaining active magnetic moments of the V ions or, more accurately, it reflects the freezing of  dimerized V ions  with a very small energy gap or with a triplet ground state.

\subsection{NMR at low temperatures, $T < 40$ K}
\begin{figure}
\includegraphics[width=0.8\linewidth]{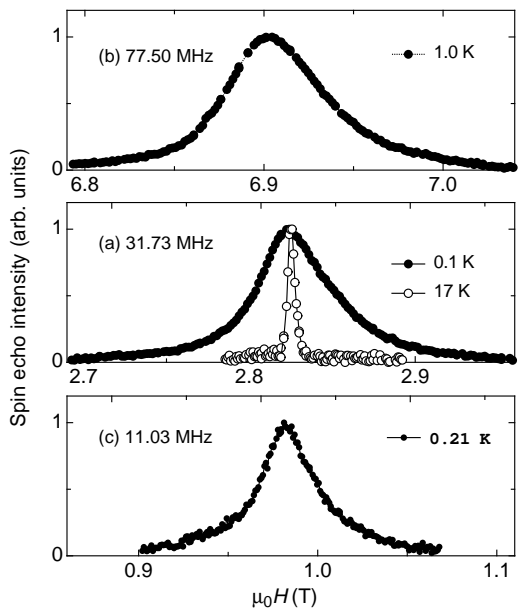} 
\caption{\label{fig9:epsart4}
Examples of the $^{23}$Na-NMR central
line of Na$_{2}$V$_{3}$O$_{7}$ for three different frequencies of
77.5, 31.73 and 11.02 MHz at very low temperatures.}
\end{figure}

In Fig.  9 we display examples of the $^{23}$Na-NMR central transition signal at very low temperatures for three fixed frequencies $\nu_{L}$ of 11.03, 31.73 and 77.5 MHz, corresponding to resonance fields $\mu_{0}H$ of 0.98, 2.825 and 6.9 T, respectively.  The comparison of the central transition signals for  $\nu_{L} = 31.73$ MHz at 17 and 0.1 K emphasizes the very large increase of the linewidth FWHM $= \Gamma_{s}$ with decreasing temperatures below 20 K (see also Ref.  8).  Since in this temperature regime, $\chi(T)$ follows a Curie type behavior with a small paramagnetic Curie temperature $\Theta_{p} \approx -2$ K, the high temperature $(T > \Theta_{p})$ NMR linewidth for a randomly oriented powder is expected to vary as $\Gamma_{s} \propto M \propto H/T$.  In order to check this prediction we plot the linewidth of the $^{23}$Na-NMR central line of Na$_{2}$V$_{3}$O$_{7}$ in Fig.  10 for three different frequencies of 77.5, 31.73 and 11.02 MHz as a function of $T/\nu_{L}$, with $\nu_{L}$ denoting the corresponding resonance frequency ($\nu_{L} \propto H$). We note that, below 40 K,  the initial increases of the linewidth with decreasing temperature fall onto the same curve for all data points, indicating that indeed $\Gamma_{s} \propto (T/H)^{-1}$.  This is emphasized by the solid line which represents a Curie type temperature dependence. This behavior breaks down at low temperatures and in Fig. 10 the onset of deviation is emphasized by the vertical arrows.  Considering that $\Gamma_{s}$ measures the width of the distribution of local susceptibilities  $\Delta \chi$, our data  suggest that $\Delta \chi$ is large at low temperatures.

\begin{figure}
\includegraphics[width=0.8\linewidth]{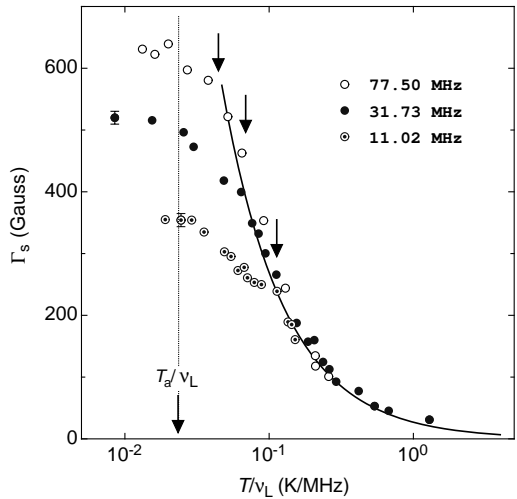} 
\caption{\label{fig10:epsart4}
 $^{23}$Na-NMR linewidth $\Gamma_{s}$ of
Na$_{2}$V$_{3}$O$_{7}$ as a function of $T/\nu_{L}$ below 40 K, with $\nu_{L}$ denoting the
resonance frequency. The vertical dotted line at $T_{a}/\nu_{L}$ separates the
temperature independent- from a strongly temperature dependent
regime. The solid line represents a Curie-type law and the arrows mark the
temperatures where $\Gamma_{s}$($T$) departs from the
high-$T$ behavior. }
\end{figure}

In Fig.  10, another feature worth to be noticed emerges. Below a constant value $T/\nu_{L} \approx 0.02$ (K/MHz), the linewidths $\Gamma_{s}$ are temperature independent, at a level that changes significantly with applied magnetic field. The vertical solid line at $T/\nu_{L} =T_{a}/\nu_{L} $ represents approximately the positions  of the maxima in $T_{1}^{-1}(T)$ at $T_{a}(H)$  reported in Ref. 8. As we already mentioned above, it is not justified to introduce a phase boundary at $T_{a}(H)$. As previously described in Ref. 8, $T_{1}^{-1}(T)$ exhibits yet another anomaly  in the form of an abrupt change of slope at $T_{b}(H)$,  distinctly  below $T_{a}(H)$. For the reader's benefit, we partly recall these features in Fig. 11. The slope change in $T_{1}^{-1}$ was interpreted as indicating the formation of a small gap in the spectrum of V spin excitations at $T_{b}(H)$. In view of lacking additional evidence from specific-heat data, we refrain from a further discussion of this feature in the present work.

\begin{figure}
\includegraphics[width=0.8\linewidth]{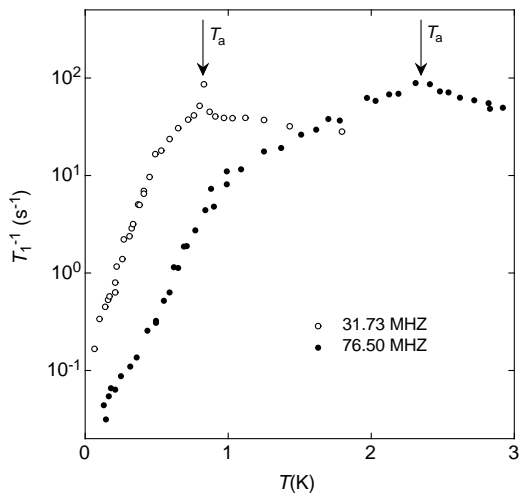} 
\caption{\label{fig11:epsart4}
Temperature dependence of the NMR spin-lattice relaxation
rate in applied magnetic fields of 2.82 and 6.81 T at temperatures
below 3 K. Prominent anomalies define $T_{a}$.}
\end{figure}

Our  $\Gamma_{s}$ data are difficult to reconcile with scenarios of either a simple type of field-induced  magnetic ordering or a spin freezing at $T_{a}(H)$.  In such cases, the linewidth at very low temperatures would be caused by internal static fields and hence would not depend on  $H$,  contrary to our observations. If, nevertheless, ``spin freezing'' is assumed to occur at $T_{a}$ then, only a minority of spins is involved in the phenomenon, as suggested by the results for the specific heat.  These particular spins are also dimerized as all the others, but either the related  singlet-triplet energy gaps are small or zero, or a triplet ground state is established. This may be plausible, but does not address the striking fact that $T_{a}$ varies linearly with $H$.  Therefore, we abandon this scenario at the outset.

\subsection{NMR at high temperatures $T > 40$ K}

\begin{figure}
\includegraphics[width=0.8\linewidth]{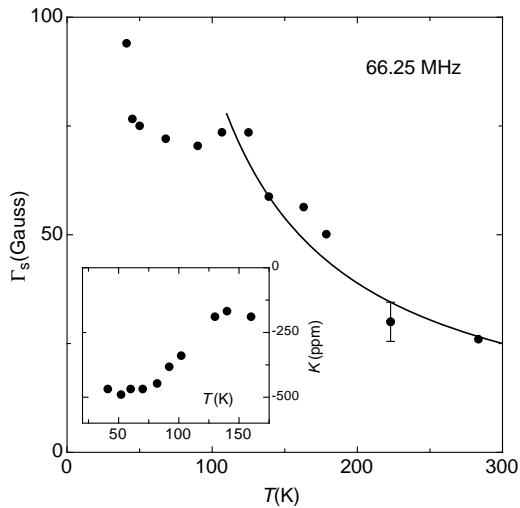} 
\caption{\label{fig12:epsart4}
$^{23}$Na-NMR linewidth ($\Gamma_{s}$) of the Zeeman
central transition of Na$_{2}$V$_{3}$O$_{7}$ as a function of $T$.
The data was measured at 66.25 MHz and the solid line emphasizes the Curie-Weiss behavior.  Inset: NMR line shift $K$ as a function of $T$
for data measured at 38.7 MHz.}
\end{figure}

Next we examine the $^{23}$Na-NMR data obtained in the temperature region above 40 K.  They  provide additional experimental evidence for the drastic change in the V $3d$ electron system around 100 K, as previously concluded
from $\chi(T)$ data in Ref. 8.  The first evidence is provided by  the temperature dependence of $\Gamma_{s}$ as shown in Fig. 12. The Curie-Weiss type enhancement with decreasing temperature, emphasized by the solid line in Fig. 12, is abruptly terminated at approximately 100 K. The plateau region between 100 and 50 K is compatible with a
loss of free ionic V moments  and hence a reduction of $n_{mag}$. The onset of yet another rapid increase of
$\Gamma_{s}$($T$), with decreasing $T$ is consistent with a very much
reduced $\Theta_{p}$ for temperatures below 40 K. 

We emphasize that at low temperatures, the width of the $^{23}$Na-NMR line is very large and is clearly dominated by the magnetism of the V moments. On the other hand, the relative shift  $K$ of the $^{23}$Na-NMR line is rather modest and the magnetism of the V moments may only be one of several factors affecting $K(T)$. Nevertheless, we note a considerable enhancement of $K(T)$ from $T \gtrsim $ 150 K to $T \lesssim $ 75 K (see inset of Fig. 12).

\begin{figure}
\includegraphics[width=0.8\linewidth]{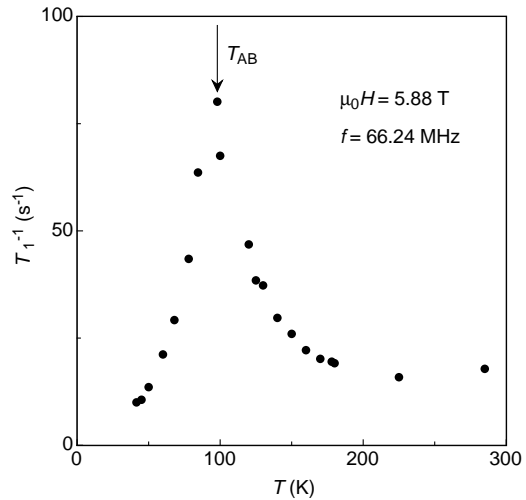} 
\caption{\label{fig13:epsart4}
Temperature dependence of the NMR spin-lattice relaxation
rate in an applied magnetic field of 5.88 T at temperatures above 40
K. A prominent maximum defines $T_{AB} $.}
\end{figure}

In Fig.  13 we show  $T_{1}^{-1}(T)$ measured in an applied field of 5.88 T. Above 200 K, $T_{1}^{-1}(T)$ is approximately
$T$-independent, as expected for a relaxation driven by the flips of paramagnetic moments.  At lower temperatures, $T_{1}^{-1}(T)$ passes through a prominent maximum  at $T_{AB}$ near 100 K. Since the relaxation is mainly due to the flips of V moments\cite{Gavilano2003}, which as mentioned above should result in a $T-$ independent relaxation, this maximum of $T_{1}^{-1}(T)$ must be associated with changes in the dynamics of the V moments. From specific-heat and magnetic-susceptibility data we infer that the most likely scenario involves the formation of V-V dimers of the V moments. In this case the loss of magnetic degrees of freedom associated with the formation of the dimers, $i.e.$, with the reduction of  $n_{mag}$, may be thought of as a gradual reduction of the fluctuation frequencies of the involved moments. This process is expected to result in a broad peak in $T_{1}^{-1}(T)$ similar to observations in precursor regions of magnetic ordering and spin freezing. The recent suggestion\cite{Ropka2003} that the reduction of  $n_{mag}$ maybe explained  by conventional crystal-field effects and spin-orbit coupling of the V$^{4+}$ ions is judged as unconvincing, because in such a scenario, the peak in $T_{1}^{-1}(T)$ is not expected.

\section{ DISCUSSION}

 \subsection{The different regions in the $H-T$ plane}

In this section we summarize the obtained experimental results for Na$_{2}$V$_{3}$O$_{7}$ and, as before, we characterize different regions in the $H-T$ plane by $n_{mag}$, which represents the concentration of those moments that dominate the increase of $\chi$ with decreasing temperature. 
 
The results for the specific heat, the NMR response (including NMR data from Ref.  8) as well as  the $ac-$ and $dc-$susceptibility, imply several characteristic temperatures that are indicated by anomalies in the temperature dependences of
\begin{enumerate}
    \item $T_{1}^{-1}$, exhibiting maxima at $T_{AB}$ and $T_{a}$,
    \item $\Gamma_{s}$,  revealing changes in the $T$ dependence at $T_{AB}$  and at $T_{a}$,
    \item $C_{m}(T)/T$, and $C_{m}(T)$, with maxima at field-dependent temperatures,
    \item $\chi_{ac}$, with a peak at $T_{max}(H=0)$ and
    \item $\chi_{dc}$, implying two different temperature regimes with different  $n_{mag}$ (see also Ref. 8).
\end{enumerate}

Well above 100 K, in a region that we denote as region $A$, all the V magnetic moments contribute to the V- based $3d$ electron susceptibility, $i.e.$, $n_{mag} = 1$. The low-temperature boundary of this conventional paramagnetic phase is marked by a very prominent and broad peak in $T_{1}^{-1}(T)$ at $T_{AB} \approx 100$ K  (see Fig. 13). Around the same temperature we note significant changes in the temperature dependences of the NMR linewidth $\Gamma_{s}$($T$), the NMR line shift $K(T)$ and $\chi(T)$. Most of these features are consistent with a gradual reduction of $n_{mag}$ upon decreasing temperature.

The observed variation of the chemical shift $K$ across 100 K (see inset of Fig. 12) most likely reflects the onset of distortions of the oxygen polyhedra surrounding the Na ions at temperatures between 75 and 125 K.  This feature and the concomitant reduction of $n_{mag}$ suggest that a dimerization of the V magnetic moments is the most likely process to account for the change of the magnetic properties from region $A$ to region $B$, the dimerized phase.  At temperatures well below $T_{AB} $, $i.e.$, between 20 and 2 K, the spin system is characterized by $n_{mag} = 1/9$.

In the same temperature regime, $i.e.$, well within region $B$, we note an excess specific heat and part of it varies linearly with $T$. Inevitably, it has to be related to excitations in the magnetic subsystem of this insulating compound. Recalling the evidence that "effectively" still $1/9$ of all the V ions act paramagnetically in this regime, it is tempting to ascribe  this $\gamma T$ term to $S = 1/2$ spinon excitations that occur in antiferromagnetic Heisenberg chains in the low-temperature limit. This interpretation does not make sense, however. In this situation\cite{Faddeev81,Takahashi73,Haldane91,McRae98}
\begin{equation}
    \gamma = C_{p}/T = 2N_{A}k_{B}^{2}/(3J) ,
    \label{eq:5}
\end{equation}
where $T \ll J/k_{B}$, and $J$ is the intra-chain exchange interaction. In view of the reduced value of $n_{mag}$, the experimental value of $\gamma$ implies that $J/k_{B} \approx 9$ K, incompatible with the request that  $T \ll J/k_{B}$ for Eq. 5 to be valid.

We emphasize once more that the anomalies in $T_{1}^{-1}(T)$ at $T_{a}$ (see Fig. 5),  the maxima of $C_{m}(T,H)$ and those of  $C_{m}(T,H)/T$ do not reflect new phase boundaries. The situation is insofar complicated that in zero magnetic field and  at temperatures below 1.5 K, region B is characterized by $n_{mag} = 1/18$.  Thus the cusp in $\chi(T)$ at $T_{max}(H \approx 0) = 0.086$ K most likely indicates a spin-freezing type phenomenon. In view of the very low concentration  of active V moments and, by consequence,  the randomness of their mutual interaction, it seems rather unlikely that a state of conventional magnetic order is adopted. As a possible scenario we suggest that the freezing at $T_{max}$ only invokes spins that occupy the very few triplet ground states that are compatible with the $P_{1}(\Delta)$ distribution of our model. The anomalies in $T_{1}^{-1}(T)$ at $T_{a}$ most likely reflect temperature-induced changes in the dynamics  of the V moments. Note that in this case the dimers of $P_{2}(\Delta)$ play only a minor role, since the vast majority of them have singlet ground states with  energy gaps relatively large compared with those of $P_{1}(\Delta)$. For applied fields of the order of 0.5 T or larger, however, the distribution of minimum excitation gaps changes considerably. The dimers  (with a triplet ground state for $\mu_{0} H = 0$ T) acquire now a gap of $2\mu_{B} H$ and play no role here. This suggests yet another phase, different from phase B and stable only at very low temperatures and in very small or zero external magnetic fields.

 \subsection{The distribution of V-V interactions }

This subsection is devoted to a discussion of our model assumptions, the calculation of the low temperature specific heat  and to an attempt to understand the complex magnetic behavior of  Na$_{2}$V$_{3}$O$_{7}$  at temperatures close to $T = 0 $ K.

First of all we explore the implications of the crucial observation that well below  $T_{AB}$ only 1/9 of the V moments contribute significantly to the Curie-Weiss type susceptibility  and that the rather small paramagnetic Curie temperature $\theta_{p} \approx -2K$ implies only weak interactions between them. We start by assuming that  8/9 of the localized V spins are strongly dimerized, $i.e.$, with rather large singlet-triplet energy gaps, and hence without much influence on $\chi(T)$ below 20 K.  The rest, 1/9, is first considered to be free. In this scenario,  $C_{st}(T) $  at low temperatures would exclusively be due to the free magnetic moments and therefore, it should be well represented by a contribution due to a collection of identical two-level systems, each with a field-dependent gap given by $\Delta =  2 \mu_{B} H$, where $H$ is the applied magnetic field.  This simple model, however, cannot account for the experimental data.The gap vanishes for $H = 0$ T, and no peak in $C_{st}(T)$ in $H = 0$ T is expected, vastly different from what we observe (see Fig. 4).  Other simple models which assume, for instance, that the spin system is formed by a collection of identical singlet-triplet units, replacing the pairs of free moments and leaving the singlet-triplet energy gap as a single-value free parameter, also fail to account for the observed $C_{m}(T)$ data.  This situation clearly suggests that the considered V spins (1/9 of the total) can neither be assumed to be free nor their local environments to be identical.  This should be contrasted with the properties of the effective spin-chirality model of odd-leg spin tubes\cite{luescher} in which the ground state is spontaneously dimerized and the effective $S = 1/2$ spins of the rings are frozen in a regular pattern\cite{footnote1}. Finally we recall that also the $\gamma T$ term in $C_{m}(T)$ needs to be accounted for.

The previous  conclusions from $\chi(T)$ and the Schottky-type features of $C_{st}(T)$ suggest that the spin system may be represented by a collection of dimerized pairs of V moments with a distribution $P(\Delta)$ of singlet-triplet energy gaps $\Delta$.  Since the situation is rather complex we proceeded in two steps. In a first step we again did not consider the $\gamma T$ term in $C_{m}(T)$, $i.e.$, we neglected the contribution to the specific heat due to the above mentioned dimers with large singlet-triplet gaps (8/9 of the total). The remaining relevant dimers were assumed to exhibit gaps with the distribution $P(\Delta)$ given in Eq. 3. The best fits to the data turned out to be consistent with rather modest values for the singlet-triplet energy gaps.  Physically, a distribution  $P(\Delta)$ of energy gaps is not unreasonable because, according to Ref. 9, a large number of different V-V couplings and, in addition,  frustration effects may be important to understand the magnetic features of Na$_{2}$V$_{3}$O$_{7}$. Our model is based on a complete dimerization of all V spins at low temperatures. Most of the dimers (8/9) are characterized by fairly large singlet-triplet energy gaps, such that their influence on $\chi(T)$ at very low temperatures may be neglected at this stage. This forces the minority of  spins, 1/9 of the total, to adopt dimerization partners from different rings, the characteristic structural element mentioned in the introduction, via weaker interactions. Therefore, due to the geometrical constraints of the structure and the expected wide range of energy characterizing the couplings between the V moments, a distribution $P(\Delta)$ of singlet-triplet energy gaps is likely to occur as the remaining  V spins dimerize. In this sense we argue that the model described in section III.a (see  Eq. 3) which is successful in reproducing the experimental $C_{st}(T,H)$ and $C_{st}(T,H)/T$ data, may be justified in the way sketched above. Naturally the same model is expected to reproduce $\chi(T)$ and we calculate the magnetic susceptibility $\chi_{1,2}(T)$ due to the dimerized spins with the  distribution of singlet-triplet gaps given by Eq. 3. The result is displayed in the main frame of Fig. 14 as the dotted line. The result of this calculation, although reflecting a Curie-Weiss type behavior of the susceptibility, is in substantial disagreement with the measured data, shown as open circles in the same figure. This is not surprising, however, because we have, as pointed out above, so far neglected the contribution  due to the strongly dimerized V spins.

Therefore, in the next step we attempted to identify the origin of the $\gamma T$ term of the specific heat and the missing contribution to $\chi(T)$. Following our previous procedure, we assumed that  also the impact of dimerized spin pairs with larger singlet-triplet gaps can be captured with a distribution of gaps, here denoted as  $P_{3}(\Delta)$. More or less by trial and error and seeking agreement with the experimental data, we established $P_{3}(\Delta)$ to be given by\begin{equation}
P_{3}(\Delta)	\hspace{0.5cm}  =  \hspace{0.5cm}  
\left\{
\begin{array}{lcl}
1.26\times10^{-3}, & \hspace{1.5cm} & -15 {\rm K}  <  \Delta/k_{B} <  250 {\rm K}  \\
1.00\times10^{-2},  &  \hspace{1.5cm}  &250 {\rm K}  <  \Delta/k_{B} <  350 {\rm K} 		\\
 0,   &  \hspace{1.5cm}  &  {\rm otherwise}.
\end{array}
\right.
\end{equation}

\begin{figure}
\includegraphics[width=0.8\linewidth]{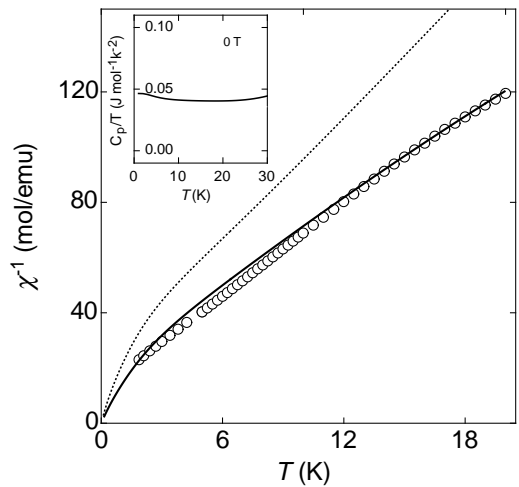} 
\caption{\label{fig14:epsart4}
Inverse susceptibility $\chi^{-1}$ vs. $T$. The circles represent the data taken from Ref. 8. The dotted line is the contribution  $\chi_{1,2}^{-1}$ due to the  distribution $P(\Delta)$ of spin singlet-triplet gaps (see text) represented in Fig. 6. The solid line is obtained by including the additional contribution to $\chi(T)$, $ \chi_{3}$,  from the distribution $P_{3}(\Delta)$ of singlet-triplet gaps (see text). The inset display the resulting contribution of $P_{3}(\Delta)$ to the specific heat.
}
\end{figure}

The prefactors have been adjusted, such that the total population (integrated area) represents $(8/9)\times(3/2)$  dimerized pairs of V moments per formula unit of Na$_{2}$V$_{3}$O$_{7}$. Note that from Eq. 6 it follows that about 3\% of the gaps $\Delta$ are of the order of 20 K or less, $i.e.$, are not ``strongly dimerized''.  With this choice of $P_{3}(\Delta)$ the model yields both, the necessary correction to $\chi_{12}(T)$ in order to reproduce fairly  well the experimental data for $\chi(T)$ (see Fig. 14), as well as the contribution to the specific heat which, in good approximation, reproduces the observed $\gamma T$ term (see the inset of Fig. 14).  We emphasize that we do not claim that the above described  $P_{3}(\Delta)$, nor $P(\Delta)$ given in Eq. 3, are the only possible or even the optimal choices for characterizing the dimers.  Nevertheless, this comparison between experimental data and model calculation gives strong support for the claim that at temperatures well below 100 K, all the V ions are dimerized. 

In summary, the complex magnetic properties of Na$_{2}$V$_{3}$O$_{7}$ may be understood by assuming that at temperatures well above 100 K the system responds paramagnetically with the localized V magnetic moments interacting with each other on a wide range of coupling energies.  At temperatures of the order of  100 K, the V moments form pairs with mostly singlet ($\Delta > 0$), but also with some triplet ($\Delta < 0$)  ground states. The comparison with experiment suggests that the singlet-triplet gaps $\Delta$ adopt values with a broad a distribution $P(\Delta)$. The total distribution $P(\Delta)$ contains at least three terms, which here are represented by two Gaussian functions, as described in Eq. 3, for 1/9 of the V moments, and an additional very broad distribution, described in Eq. 6, which characterizes the rest of the V moments (8/9). The evidence that leads to the claim that the low-temperature magnetic features are dominated by only 1/9 of the moments is most clearly provided by the respective entropy release shown in Fig. 7.

The model calculation captures the variation of the magnetic properties over a broad temperature range around and below $T_{AB}$. The peak in $T_{1}^{-1}(T)$  at $T_{AB}$ reveals that the dimerization process must be accompanied by  dramatic changes in the dynamics of the V moments. Further changes in the dynamics of the V moments, this time involving a slowing down of the fluctuations between the occupied energy levels of the dimers, seem to occur at much lower temperatures, of the order of $T_{a}$, as judged by the anomalies in $T_{1}^{-1}(T)$ shown in Fig. 11.  In a first approximation we argue that in this regime, $T_{1}^{-1}(T)$ is basically determined by transitions between spin levels of individual dimers with low energy gaps. The relaxation due to a collection of identical V dimers is given by\cite{Abragam}
\begin{equation}
   T_{1}^{-1} = \frac{A\tau}{(1+\omega^{2}\tau^{2})} ,
      \label{eq:7}
\end{equation}
with $\tau$ as the correlation time of the dimer spin transitions, $\omega = 2\pi\nu_{L}$ with $\nu_{L}  \propto  H$ as the Larmor frequency and $A$ is a constant which includes the hyperfine-field coupling between the considered dimers and the probed nuclei. Assuming that the $T_{1}^{-1}$ variation is predominantly due to a variation of $\tau$, Eq. 7 provides a maximum of $T_{1}^{-1}(\tau)$ for $\omega\tau = 1$. It also seems reasonable to assume that the transitions are thermally activated. In the simplest case, the transitions are activated  over a single barrier $U$ and $1/\tau \propto \mathrm{exp}(-U/k_{B}T)$. In case of a distribution $f(U)$ of barriers $U$, an effective correlation time  $\tau_{eff}$ is given by 
\begin{equation}
   \frac{1}{\tau_{eff}} \propto  \int \exp \left( -\frac{U}{k_{B}T} \right)f(U) dU . 
     \label{eq:8}
\end{equation}
If $f(U)$ is assumed to be constant for the relevant values of $U$, this reduces to
\begin{equation}
   \frac{1}{\tau_{eff}} \propto  \int \exp \left( -\frac{U}{k_{B}T} \right) dU \propto T .
     \label{eq:9}
\end{equation}
Eqs. 7 and 9 explain a strong increase of $T_{1}^{-1}$ with decreasing temperature in the limit 
$\omega \tau_{eff} << 1$ where  $T_{1}^{-1} \approx  A\tau \propto T^{-1}$. Likewise the decrease of  $T_{1}^{-1}$ beyond the maximum  is also consistent with our assumptions, because  $T_{1}^{-1} \approx  A/(\omega^{2}\tau)  \propto T$ for $\omega\tau_{eff}>>1$. Since the maximum in $T_{1}^{-1}(T)$ is tied to the condition $\omega\tau = 1$, it follows that $T_{a} \equiv T_{max}  \propto 1/\tau \propto \omega \propto H$, as observed experimentally. We are well aware that Eq. 7 only crudely captures the characteristics of our spin system.  Without more detailed information on the spin system, however, we cannot offer a more sophisticated description of the dynamics of the dimers. Nevertheless, the fact that the data is consistent with $\tau_{eff} \propto 1/T$ suggests that the spin dynamics is  rapidly slowing down with decreasing temperatures. As may be seen in Fig. 5, $T_{max} \approx 0.2 $ K for $\mu_{0}H \approx 1$ T. This field value corresponds to a frequency of 11.02 MHz\cite{Gavilano2003}, a rather low value. The extrapolation of $T_{a}(H) \equiv T_{max}(H)$ shown in Fig. 5  suggests again that the system is close to a QCP, as previously claimed\cite{Gavilano2003}.

Finally we point out that one aspect of our data, the temperature independent  but field dependent NMR linewidth $\Gamma_{s}$ at low temperatures (see Fig. 10), is not explained by our model.  Commonly $\Gamma_{s}(T,H)$ is proportional to the distribution of local fields $\Delta h_{loc}$ at the probed nuclei. As demonstrated in Fig. 10, this is also true in our case at temperatures exceeding $T_{a}(H)$ and consistent with our model which implies a monotonic increase of the total magnetization  $M$ with decreasing temperature. In this sense, the abrupt change of behavior of $\Gamma_{s}$ at $T_{a}/\nu_{L}$ is unexpected, indicating that our model misses an important ingredient which is effective only at low temperatures.  The relevant dimers (with small gaps) in Na$_{2}$V$_{3}$O$_{7}$  are expected to be rather extended in real space. The implicit model assumption of non interacting dimers may thus be a rather crude approximation. A more detailed knowledge of the couplings between  the V moments seems to be  needed for a self consistent interpretation all the experimental data of  Na$_{2}$V$_{3}$O$_{7}$.

\section{CONCLUSION}

The magnetic properties of Na$_{2}$V$_{3}$O$_{7}$ are qualitatively different in different regions in the $H-T$ plane. These differences may be thought of as the result of a gradual "quenching" of the localized V moments with decreasing temperatures below approximately 100 K. But from the analysis of the specific heat we conclude that this "quenching" reflects a process of dimerization at temperatures of the order of 100 K, resulting in a distribution $P(\Delta)$  of singlet-triplet energy gaps $\Delta$. For 1D systems, a simple dimerization generally leads to observable changes in the crystal structure. Such changes have been searched for, but were not observed in optical studies\cite{choi2003}.  From our results it is suggested that the $V-V$ couplings involve a broad range of interactions and given the unusual structural features of Na$_{2}$V$_{3}$O$_{7}$, which may include some randomness, it is not clear that changes in the structure that could be detected in scattering and/or optical experiments should be expected here. We note, however,  that the temperature induced variation of the $^{23}$Na chemical shift shown in the inset of Fig. 12, may well be due to subtle structural changes accompanying the  compensation  of the moments.

Anomalies in $T_{1}^{-1}(T,H)$ and a cusp in $\chi(T)$ at 0.086 K reveal that Na$_{2}$V$_{3}$O$_{7}$ is close to a quantum critical point, as previously claimed. Finally the broad distribution of couplings suggests that the relevant low-energy effective magnetic model should contain some level of randomness\cite{Dasgupta94,Fisher94,Sigrist94}. Given the complexity  of the V network,  it is at present impossible to say whether these properties, including the behavior of $\Gamma_{s}(T,H)$ at low temperatures, may  consistently be explained with a purely magnetic model with some coupling to the lattice, or whether some extrinsic disorder, for instance due to some missing  Na$^{+1}$ ions in the structure, has to be taken into account.

\section{ACKNOWLEDGEMENTS}

This work was financially supported by the Schweizerische Nationalfonds zur F\"{o}rderung der Wissenschaftlichen Forschung (SNF). We have benefitted from a number of instructive discussions with M. Sigrist and V. Mazurenko. The study also profited from support of the NCCR program MaNEP of the SNF.




\begin{thebibliography}{99}
\bibitem{footnote1} Note that the spin-chirality model of Ref. \cite{luescher} is also inconsistent with our specific-heat data since, for appropriate parameters, it predicts a low-temperature narrow peak due to chirality degrees of freedom, which  we do not observe.
\end{thebibliography}
\end{document}